\documentclass[twoside,leqno,twocolumn]{article}

\usepackage[letterpaper]{geometry}
\usepackage{csquotes}
\usepackage{ltexpprt}
\usepackage{balance}
\usepackage{graphicx,wrapfig,xcolor}
\usepackage{amsmath,amsfonts,amssymb,mathtools}
\newcommand{\R}{\mathbb{R}}
\newtheorem{mydef}{\textbf{Definition}}
\usepackage{balance}
\usepackage{algorithm}%
\usepackage{algpseudocode}%
\usepackage{makecell}
\usepackage{caption}
\usepackage{subcaption}
\usepackage{booktabs}
\usepackage{tabularx} 
\usepackage{graphicx}
\usepackage{multirow}
\usepackage{hyperref}

\begin{document}

\title{Dissecting Ethereum Blockchain Analytics: What We Learn from  Topology and Geometry of Ethereum Graph}


\author{Yitao ~Li\thanks{Purdue University, USA.}
\and Umar ~Islambekov \thanks{Bowling Green State University, USA.}
\and Cuneyt Akcora  \thanks{University of Manitoba, Canada.}
\and Ekaterina Smirnova \thanks{Virginia Commonwealth University, USA.}
\and Yulia ~R. Gel \thanks{University of Texas at Dallas, USA.}
\and Murat Kantarcioglu\footnotemark[5]}

\date{}

\maketitle


\fancyfoot[R]{\scriptsize{Copyright \textcopyright\ 2020 by SIAM SDM 2020\\
Unauthorized reproduction of this article is prohibited}}





\begin{abstract} \small\baselineskip=9pt Blockchain technology and, in particular, blockchain-based cryptocurrencies offer us information that
has never been seen before in the financial world. In contrast to fiat currencies, {\it all} transactions of crypto-currencies and crypto-tokens are permanently recorded on distributed ledgers and are publicly available. As a result, this allows us to construct a transaction graph and to assess not only its organization but to glean relationships between transaction graph properties and crypto price dynamics. The ultimate goal of this paper is to facilitate our understanding on horizons and limitations of what can be learned on crypto-tokens from local topology and geometry of the Ethereum transaction network whose even global network properties remain scarcely explored. By introducing novel tools based on topological data analysis and functional data depth into Blockchain Data Analytics, we show that Ethereum network (one of the most popular blockchains for creating new crypto-tokens) can provide critical insights on price strikes  
of crypto-tokens that are otherwise largely inaccessible with conventional data sources and traditional analytic methods.
\end{abstract}

\section{Introduction}

Past few years marked the beginning of a new era of technology -- the era of Blockchain. Blockchain has already revolutionized many fields, from e-payments to digital asset ownership management. Undoubtedly, one of the primary magnets of the Blockchain craze is to take advantage of the unprecedented opportunities to invest (and to lose!) in the largely unregulated crypto-markets via various forms of digital instruments such cryptocurrencies and crypto-tokens. Recent sharp soars and miraculous comebacks of crypto-assets only continue to further heat the investment mania due to unprecedented chances to make a quick, and strikingly high profit, which in turn goes hand in hand with high investment risk. 
Naturally, one of the most momentous questions nowadays are {\it whether}, {\it how} and {\it to what extent} we can forecast trading dynamics of crypto-currencies and tokens.

Despite many novel analytic challenges associated with blockchain-based financial instruments, the crypto-market offers us highly informative and novel data that have never been available before due to data protection and privacy policies in banking -- that is, all transactions are permanently recorded and publicly available. This in turn allows us, for the first time in the history of finance, to construct a global {\it transaction graph} and relate its properties to price dynamics. The intuition behind this approach is multi-fold.
 
First, as various patterns of retail shopper activity provide a foundation for assessment of the current state of the economy and form the basis of many 
economic indicators~\cite{reinsdorf2009review}, it is natural to hypothesize that patterns of the transaction graph may also offer a glimpse into crypto-market health.

Second, in contrast to retail shopper data that are both heavily aggregated and delayed in time, information on the transaction graph is available in real time and on a transaction-level basis. 

Third, availability of the transaction graph allows to study speculative and even malicious behavior of crypto-assets' users which, as often happens in the financial world, involves multiple players -- such behavior remains 
largely inaccessible with conventional analytics and requires tools of complex network inference.   

We show that geometry and local topology of the transaction graph contains a wealth of information on crypto-token market, ranging from price prediction and price anomalies to hidden co-movement of multiple instruments.

In contrast, we find that both conventional variables of financial time series and \textit{global network features} of the transaction graph are not capable to glean a deeper insight into crypto price dynamics.
 
\begin{figure}[h]
    \centering
    \includegraphics[width=0.30\textwidth]{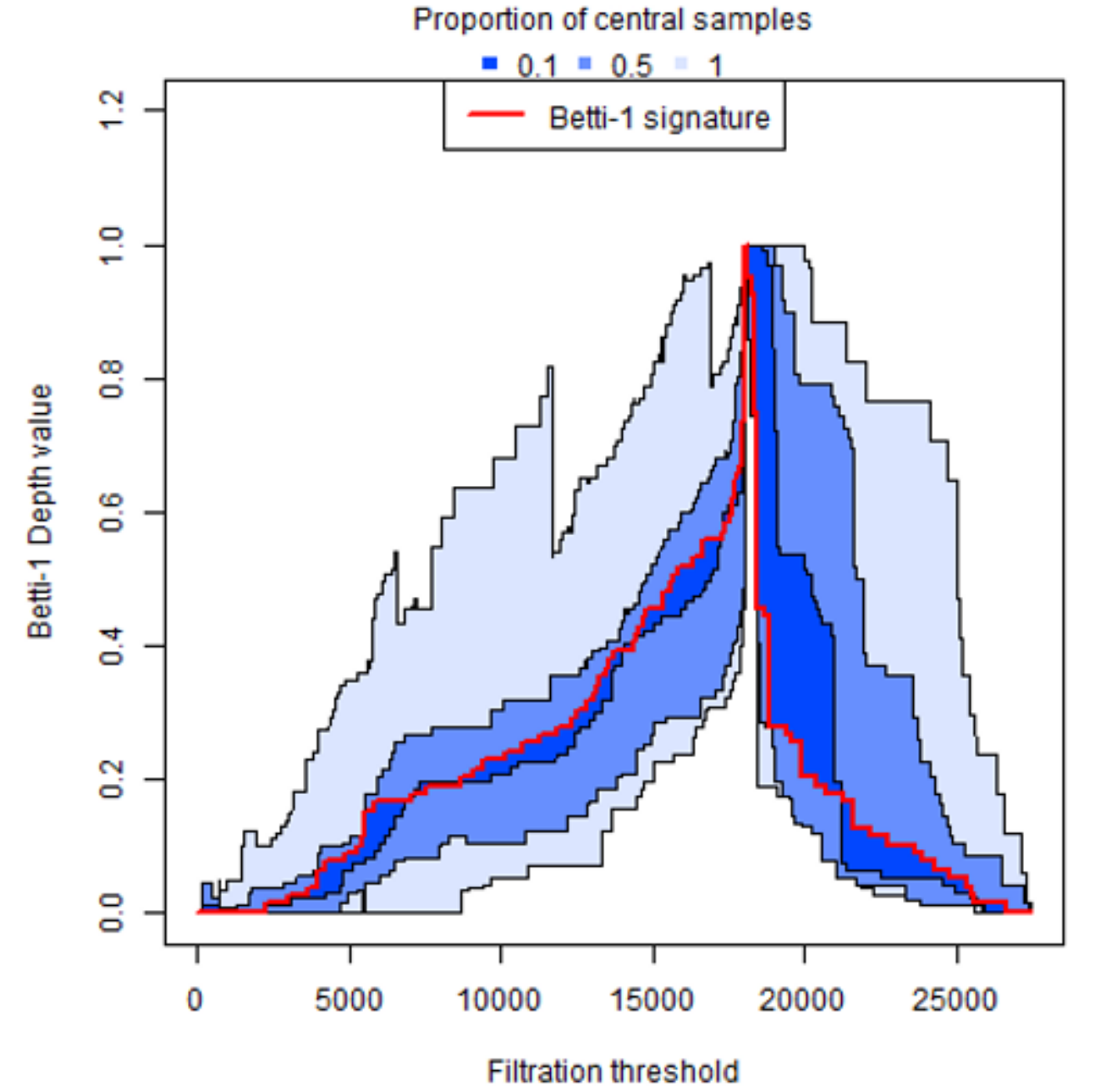}
    \caption{Betti pivot of the token Powerledger (shown in red).}
    \label{fig:tokensignatures}
\vspace*{-0.6cm}    
\end{figure}

{\bf Why Ethereum?} Ethereum is one of the {\it the most popular} blockchain platforms. It allows creating smart contracts and, hence, enables everyone to create a crypto-asset on it. Ethereum tokens are sold through Initial Coin Offerings (ICOs). Such token ICOs have already enabled many start-ups and organizations to raise capital by selling digital coins which allow recipients to use a promised service if and when available. Only in 2018 and 2019, Ethereum ICOs raised billions of dollars. As such, analysis of Ethereum networks might be even more acute than cryptocurrency price prediction. However, despite this high token activity, network structure of Ethereum transaction graph
remains largely understudied
\cite{chan2017ethereum, ferretti2019ethereum}. 
Furthermore, to the best of our knowledge, there exist no studies of Ethereum that link crypto-token price analytics with the underlying Ethereum transaction graph.

{\bf Why Not Simpler Methods?}  
Ethereum data poses several challenges. First, the transaction graph is very sparse and dynamic. Nodes (i.e., account addresses) appear and disappear (i.e., no future transaction) daily, while the number of transactions widely fluctuates across days.  Hence, conventional graph analytic tools such as global clustering coefficient and $k$-core~\cite{dorogovtsev2006k} analysis may not be feasible indicators of token activity. Second, crypto-tokens may exhibit very different responses to external positive and negative shocks, and the 
 signal on such a response to shocks which is contained in the crypto-token price and global network features, is weakened due to aggregation.     
 These challenges require development of novel \textit{robust graph theoretic approaches that are suitable for analysis of time-varying, highly irregular, and very sparse networks}.   
  
{\bf Proposed Approach.}  We address the above-mentioned challenges by introducing the arsenal of topological data analysis (TDA) tools into Ethereum data analytics.
TDA allows us to systematically and robustly assess a local geometric and topological structure of the Ethereum transaction graph.
Our approach is based on the premise that any abnormal situation, for instance price anomaly, viewed as a response to a negative or positive shock (e.g., announcement of a new crypto-currency regulation) is likely to be reflected in the underlying topology and geometry of a transaction graph. To study the local network geometry and topology of the Ethereum transaction graph, we blend concepts from algebraic topology and functional data analysis.  

The important methodological distinction of our new approach is that while TDA has been applied before to financial time series, including time series of cryptocurrencies~\cite{gidea2018topological}, {\it TDA has  never been yet applied to complex networks of financial transactions on account-based blockchains
such as Ethereum}. Moreover, to the best of our knowledge, the only other paper discussing utility of TDA on financial networks, including {\it both traditional finance and blockchain}, 
is our earlier study of Bitcoin graph~\cite{abayChainnet18} which belongs to the   unspent  transaction  output  (UTXO)  based blockchains. Since UTXO based blockchain graphs have transactions with multiple inputs and outputs, the techniques developed for UTXO blockchains cannot be directly applied to account-based blockchains.

As such, the importance of our methodology and findings can be summarized as follows:

\medskip
\noindent $\bullet$ 
We offer a novel perspective to risk analysis of crypto-assets, particularly, Ethereum tokens, by dissecting hidden linkages between the token price dynamics and local geometry of the Ethereum transaction graph. While the paper focuses on blockchain data analytics, the proposed novel methodology to risk analysis based on geometry and topology of the transaction graph is applicable beyond crypto instruments. For instance, subject to data availability on transactions and other financial interactions, the proposed analytic tools can be used for analysis of systemic risk in interbank networks as well as for optimizing strategies in algorithmic stock trading.

\medskip
\noindent $\bullet$ We propose a new measure of the most illustrative, or \textquote{normal} behavior on the Ethereum transaction network: a {\bf Betti pivot}. Betti pivots, based on analysis of network persistent homology and functional data depth, 
allow us to quantify and visually assess differences between normal and anomalous transaction activity, as we show in  Fig.~\ref{fig:tokensignatures} for the PowerLedger  token.  

\medskip
\noindent $\bullet$ We develop an innovative filtering approach that significantly reduces the  (prohibitively high) computational costs of TDA. We report the first results where TDA tools can be adopted in large networks while preserving the performance. 
 
\medskip
\noindent $\bullet$ We report the first results for crypto-token price anomaly prediction, and show that token networks are likely to contain adequate information to develop
arbitrage trading strategies in the real world. As the crypto-token ICOs have reached \$12B in the first half of 2018~\cite{forbes}, our prediction results have important real-life implications for start-up funding.

\section{Related Work}
 
 We outline four relevant research areas: Ethereum graph analysis, Blockchain price prediction and anomaly detection, as well as TDA.
 
 \medskip
\noindent\textbf{Ethereum graph analysis.} Differing from crypto-currencies (e.g., Bitcoin) 
 where each transaction can have multiple inputs and outputs~\cite{ron2013quantitative}, Ethereum transactions 
 transfer ether or tokens from one address to another. As such, Ethereum lends itself to  traditional network analysis. 
 For instance,
 \cite{anoaica2018quantitative} studied empirical properties of Ethereum and 
 \cite{somin2018network} explored 
 token networks,
 in terms of degree distribution, power laws and clustering. 
 However,  there are yet no results that 
employ network tools for Ethereum price analytics. 

\medskip
 \noindent\textbf{Cryptocurrency price prediction}. Analyzing transactions and addresses to track the Bitcoin economy has become an important research direction. 
 A time series prediction approach by~\cite{mcnally2018predicting} uses a Bayesian optimized RNN and LSTM network with varying degrees of success. Blockchain features, such as average transaction amount, 
 are also shown to exhibit mixed performance for cryptocurrency price forecasting~\cite{greaves2015using}. Various blockchain graph characteristics, such as average degree, can be used as prediction features.
 Recently, \cite{akcora2018forecasting} employed blockchain motifs, termed chainlets, as features to predict Bitcoin price.  
 
However, all the mentioned approaches are carried out to track a single cryptocurrency. In contrast, our goal is to track multiple cryptoassets at the same time.  
 
 \medskip
\noindent\textbf{Blockchain anomaly detection.}
Blockchain addresses can be linked to identify people behind suspicious transaction patterns in cryptocurrencies~\cite{spagnuolo2014bitiodine}. The pattern is usually defined as a repeating shape that involves moving coins from a (black) address to an online exchange, where the coins can be cashed out without being confiscated by authorities. The black address that starts the transaction chain may be related to money laundering~\cite{moser2013inquiry}, blackmailing~\cite{Egretpaper} and  ransomware payments~\cite{huang2018tracking}. There exists ample evidence 
of these anomalies in the transaction network~\cite{bogner2017seeing}. 
A more recent approach found anomalies in Bitcoin price by linking addresses to transactions in time~\cite{griffin2018bitcoin}. In contrast, we do not assume any prior knowledge about pattern shapes or addresses; our unsupervised data depth approach tracks token networks for price anomalies. 
 
\medskip
\noindent \textbf{Topological Data Analysis.} TDA is an emerging field at the interface of algebraic topology, statistics, and computer science. The rationale is that the observed data are sampled from some metric space and the underlying unknown geometric structure of this space is lost due to sampling. The key idea is to recover the lost underlying topology~\cite{Wasserman:2018}. Persistent homology (PH) is one of the tools to characterize a topological data structure under varying scales of dissimilarity.
The most widely used topological summaries of persistent features are the Betti numbers, barcode plots, persistent diagrams, and persistent landscapes~\cite{Ghrist:2008}. 
However, barcode plots and persistent diagrams cannot be easily used 
in machine learning models~\cite{Bubenik:2015}.  
Differing from these approaches, we propose Betti pivots, which can be directly integrated with functional data analysis tools.

\section{Methodology}
\label{sec:methodology}

{\noindent
{\begin{minipage}{1\linewidth}
\textbf{Problem Statement:} 
Given the transaction network of an Ethereum token and time series of the token prices in fiat currency, predict whether the token absolute price return $ \mbox{R}_t = ({{Price}_{t} -  {Price}_{t-1}})/({{Price}_{t-1}})$
will change more than $|\delta|>0$, in the next $h$ days. Furthermore, identify the maximum horizon value $h$ such that the prediction accuracy is at least $\rho$. That is, the ultimate goal is to predict whether a window of the next $h$ days, $h>0$, will contain a price return anomaly. In return, an informed and reliable answer to this question allows to optimize investment strategies in algorithmic trading, and higher $h$ is preferred. 
\end{minipage}}
}
 
The key idea behind our approach is the following: first, armed with TDA,  extract multi-resolution topological summaries of the Ethereum network and then  incorporate the resulting geometric information into analysis of token prices. As the primary TDA methodological engine, we employ the tool of persistent homology due to its flexibility in integration with machine learning models.
 
  We start by detailing persistent homology and associated topological summaries.

\subsection{Persistent Homology and a New Look
at Its Summaries via Functional Data Analysis -- Betti Limits} 
\label{Background}
 Let $G=(V,E,\omega)$ be a weighted graph, where $V$ and $E$ are the set of nodes and edges, respectively; $\omega :E \rightarrow \R \cup\{\infty\}$ is a weight function encoding similarity between two nodes connected by an edge. 
To account for dissimilarity between two disconnected nodes, we introduce the weight 
$\tilde{ \omega }:V \times V \rightarrow \R\cup\{\infty\}$
 
\begin{equation*} 
\tilde{\omega}_{uv}=
\left\{
        \begin{array}{ll}
            \omega_{uv} & (u,v)\in E, \\
            \infty & (u,v) \notin E. 
        \end{array}
    \right\}
\end{equation*}
where
\begin{equation*} 
\omega_{uv} =\Big[1+\alpha\cdot\frac{A_{uv}-A_{min}}{A_{max}-A_{min}}\Big]^{-1}.
\end{equation*}
 
Here $A_{uv}$ is the amount of transferred tokens by transactions between nodes $u$ and $v$; $A_{min}$ and $A_{max}$ are the smallest and largest transaction amounts, respectively. That is, the larger the transferred amount, the smaller the inter-nodal dissimilarity. 
We set $\alpha=9$ to map weights to the interval $[0.1, 1]$.

The most important aspect of Persistent Homology (PH) is that it allows us to analyze data at multiple spatial resolutions in a unified way, bypassing a subjective selection of the dissimilarity parameter or searching for its optimal value. However, to be able to extract topological information from a point cloud, it needs to be equipped with a structure of a topological space. In the context of PH, this is commonly achieved by constructing an abstract simplicial complex on the top of data points.

\begin{mydef}[Abstract simplicial complex]
		Let $X$ be a discrete set. An abstract simplicial complex is a collection $\mathcal{C}$ of finite subsets of $X$ such that if $\sigma\in\mathcal{C}$ then $\tau\in\mathcal{C}$ for all $\tau\subseteq\sigma$. If $|\sigma|=p+1$, then $\sigma$ is called a \emph{$p$-simplex}.
\end{mydef}
Intuitively, a simplicial complex can be viewed as a higher dimensional generalization of graphs which represents a structure consisting of points, edges, triangles and their higher order counterparts. 
\emph{Vietoris-Rips} is a widely used  simplicial complex due
to  its easy construction and fast computational implementation~\cite{Carlsson:2009}.

\begin{mydef}[Vietoris-Rips complex]
		Let $X$ be a discrete set in some metric space. A Vietoris-Rips complex on $X$ at dissimilarity scale $\epsilon\geq 0$, denoted by $VR_\epsilon$, is an abstract simplicial complex whose $p$-simplices, $p=0,\ldots,d$, consist of points which are pairwise within distance of $\epsilon$. Here, $d$ is called the dimension of the complex. 		
\end{mydef}
Remarkably, simplicial complexes can not only be regarded as topological spaces from which topological information is derived, but also as combinatorial objects which are convenient for computational purposes. Hence, this dual nature of simplicial complexes turns the task of extracting topological information into a computationally feasible combinatorial problem \cite{TDAintro}.

Now, we fix a sequence of scale resolutions  $\epsilon_1<\epsilon_2<\ldots<\epsilon_n$ and form a chain of nested  VR complexes called a \textit{(finite) VR filtration} 
 $VR_{\epsilon_1} \subseteq VR_{\epsilon_2} \subseteq \ldots \subseteq VR_{\epsilon_n}$,
where $VR_{\epsilon_k}$, $k=1,\ldots,n$, is a VR complex on $V$ 
such that $VR_{\epsilon_k}=\bigl\{\sigma \subset V| \tilde{\omega}_{uv}\leq \epsilon_k, \forall u,v\in\sigma\bigr\}$.

Armed with the VR filtration, we now get a formal multi-resolution glimpse into the Ethereum network topology and geometry and track topological features that appear and later disappear as the scale parameter increases. Evolution of such topological features sheds light on organization of the Ethereum transaction network. That is, we can expect that features with a   
longer lifespan, i.e. {\it persistent features}, have a higher role
in explaining functionality of the Ethereum network than features with a shorter lifespan. These short term features are regarded as {\it topological noise}.  
Persistent features are instrumental for distinguishing anomalous dynamics in token transaction activities.  
We extract descriptors of such topological features at a multi level in the form of sequences of \textit{Betti numbers}.
  
\begin{mydef}[Betti number]
Betti-$p$ number of a simplicial complex $\mathcal{C}$ of dimension $d$, denoted by $\beta_p(\mathcal{C})$, is defined as the rank of the $p$-th homology group of $\mathcal{C}$,  
$p=0,1,2,\ldots,d$. 
\end{mydef}
Fortunately, for applied data analysis Betti-$p$ number has a simpler practical interpretation, i.e.
Betti-$0$ is the number of connected components,
Betti-$1$ is the number of loops (or holes), Betti-$2$ is the number of voids (or cavities), etc.
 
In this paper, we consider features up to dimension 2 and take $\mathcal{C}$ to be a VR complex. Following the PH methodology, we compute sequences of Betti numbers of a chain of nested VR complexes 
and thereby track the counts of different topological features at increasing scales of complexity. 
Note that the resulting topological descriptors in the form of Betti numbers over a VR filtration depend on $\epsilon_k$ 
and are intrinsically infinite dimensional. As such, an intuitive approach to analyze their dynamic properties is via functional data analysis (FDA)~\cite{ramsay2004functional, wang2016functional}. 
In this context, \textit{we introduce a novel concept of} \textbf{Betti limits} \textit{which relates these counts to the scale parameter viewed as continuum}.

 \begin{mydef}[Betti limit]\label{Betti_function}
Let $\{\mathcal{C}_{\epsilon_k}\}_{k=0}^n$ be a filtration of simplicial complexes where $\{\epsilon_k\}_{k=0}^n$ is an increasing sequence of scales such that $\epsilon_0=0$ and $\epsilon_n=L$ for some $L>0$. 
Then, the Betti-$p$ limit $\mathcal{B}_p: [0,L]\rightarrow \mathbb{Z}\cup\{0\}$, $p=0,\ldots,d$, is defined as
$$
\mathcal{B}_p (\epsilon) 
=\lim_{\max \Delta \epsilon_k\rightarrow 0} \beta_p(\mathcal{C}_{k^*})
$$
where the max is taken over all $k=0,\ldots,n$,  $\Delta \epsilon_k=\epsilon_k-\epsilon_{k-1}$ and $k^*\text{ is the index such that } \epsilon\in[\epsilon_{k^*-1},\epsilon_{k^*})$

\end{mydef}

The Betti limits can be regarded as functional summary statistics of the network's topological structure and
offer multi-fold benefits.
First, the Betti limits provide a systematic linkage with the tools of functional data analysis (FDA). For instance,
underlying nonlinear dynamics of the Betti limits can be then assessed with derivatives and associated manifold learning and empirical differential equations. In turn,
relative positions of individual trajectories of the Betti limits can be quantified using a concept of \textit{functional data depth}. Furthermore, Betti limits
can be viewed as {\it generalized} descriptors of network topology for a class of continuous latent space models, particularly, including distance models and graphons~\cite{caron2017sparse, smith2017geometry}. We leave this more fundamental mathematical hypothesis on characterizing geometry of the continuous latent space network models via Betti limits for future research.

\subsection{Functional Data Depth of Betti Limits}
 Let $\{(G_t,\tilde{\omega}_t)\}_{t=1}^T$ be a time series of weighted graphs and $\{\mathcal{B}_{p,t}\}_{t=1}^T$ 
be the associated sequence of Betti limits. 
To assess which topological
descriptors (or equivalently which transaction networks) signal towards anomalous patterns relative to others, we employ the notion of {\it data depth}.  

\begin{mydef}[Data Depth] 
Informally, data depth is a function that measures how closely a given multivariate observation is located to the ``center'' of the observed point cloud. That is, data depth extends the concept of quantiles from univariate to multivariate distributions. Formally, let $\mathcal{F}$ be a set of probability distributions on a Banach space $\mathcal{X}$ (e.g., $\mathcal{X}=\R^n$). A data depth is a function $D:\mathcal{X}\times\mathcal{F}\rightarrow [0,1]$ such that $D(\cdot|F)$ is a center-outward ordering of elements of $\mathcal{X}$ with respect to $F$. (Here, by a center-outward ordering, we mean that each element of $\mathcal{X}$ is assigned a score from 0 to 1 such that a higher score implies that the element is more centrally located within a point cloud and a lower score implies that the element is likely to be an outlier in respect to the remaining elements.) The depth of $y\in\mathcal{X}$ with respect to $\{y_i\}_{i=1}^m\subseteq \mathcal{X}$, denoted by $D(y|y_1,\ldots,y_m)$, is defined as $D(y|\hat{F}_m)$, where $\hat{F}_m$ is the empirical distribution of $\{y_i\}_{i=1}^m$.
\end{mydef}

Since we focus on Betti limits, we resort to functional data depths (i.e., where $\mathcal{X}$ is a space of functions). Among such functional depths, the modified band depth (MBD)~\cite{MBD} is particularly well-suited for detecting anomalies as MBD accounts for both the shape and magnitude of the function graphs. In addition, MBD is robust and enjoys fast computational implementation. However, our framework is sufficiently general and can be integrated with any functional data depth function.

\begin{mydef}[MBD]
Let $B(I)$ be the Banach space of bounded functions on interval $I$ and $\lambda$ be the Lebesgue measure. Given $\mathcal{Y}=\{y_1,y_2,\ldots,y_m\}\subseteq B(I)$. The MBD of $y \in B(I)$ with respect to $\mathcal{Y}$ is 
 $$MBD(y|\mathcal{Y})= {m \choose 2}^{-1}\lambda(I)^{-1} \sum_{1\leq i_1\leq i_2\leq m} \lambda(A(y;y_{i_1},y_{i_2}))$$
where  $A(y;y_{i_1},y_{i_2})=\{x\in I:\underset{r=i_1,i_2}{min}y_r(x)\leq y(x) \leq \underset{r=i_1,i_2}{max}y_r(x)\}$
\end{mydef}
Intuitively, $MBD(y|\mathcal{Y})$ measures the extent to which the graph of a function $y$ lies within the bands determined by the graphs of all possible pairs from $\mathcal{Y}=\{y_1,y_2,\ldots,y_m\}$. MBD enables us to order a set of functions in $[0,1]$-scale, where the depth values closest to zero and one correspond to the most anomalous and central functions, respectively. 
We introduce a concept of \textit{Betti pivots} which is defined as the deepest or most central Betti limit. 
\begin{mydef}[Betti pivot]
\vspace*{-0.5em}
For a given collection of Betti limits $\{\mathcal{B}_{p,t_1},\mathcal{B}_{p,t_2},\ldots,\mathcal{B}_{p,t_m}\}$, their Betti pivot is defined as
$$\mathcal{B}_p^s=\underset{\mathcal{B}_{p,t}\in\{\mathcal{B}_{p,t_1},\ldots,\mathcal{B}_{p,t_m}\}}{arg max}MBD(\mathcal{B}_{p,t}|\mathcal{B}_{p,t_1},\ldots,\mathcal{B}_{p,t_m})
$$
\end{mydef}
To measure how the Betti limits change over time and compare with the ones prior to them, we calculate the MBD depth of each day's Betti limit with respect to those of the past $w$ days. We introduce a notion of {\it rolling depth (RD) on Betti limits}
\begin{eqnarray}
\label{RollDepth}
\hspace*{1cm} RD_w(\mathcal{B}_{p,t}):= MBD(\mathcal{B}_{p,t}| \mathcal{B}_{p,t},
\ldots\mathcal{B}_{p,t-w+1}).
\end{eqnarray}

Note that RD $\in (0,1)$ and shows the position of the Betti limit on any given day $t$, relative to the past $w$ days. In turn, the Betti pivot yields the most central, or the "baseline" behavior of Betti limits over a subset of days $t_1,\ldots,t_m$. The concept of RD echoes the rolling window approaches used to detect signals of short and long term trends in algorithmic trading and to construct stock price indicators such as percentage price oscillator and moving average convergence divergence~\cite{Zakamulin2017}.

\vspace*{-0.5em}
\subsection{Anomaly Detection with Topological Features}
\label{sec:methanomaly}
  
We label a day $t$ as anomalous in Ethereum token trading,  if there is a price shock on day $t$, that is, if $|\mbox{R}_t|\geq \delta$, where $\delta>0$ is a trader-defined threshold (i.e., magnitude of a price shock) (see Problem Statement in Section~\ref{sec:methodology}). We combine new graph topological features with traditional network summaries and build one predictive model for each token. We then examine model performance for different prediction horizons $h >0$.

Our token-based price anomaly detection methodology for Ethereum crypto-tokens problem is summarized as follows.  For  each day, $t$, with available token data, we calculate the binary flag variable with values equal to $true$ if price strike in terms of the absolute token price return ($|\mbox{R}_t| \geq \delta$), has been detected in at least one of the next $h$ days (i.e., days $t+1, t+2, \dots, t+h$) and $false$ otherwise. Here, $t = 1, \dots, T_k$ is the set of historical dates  for which we have the $k$-th token data. For day $t$, we compute  the  token's normalized open price,
$\mbox{PN}_t = Price_t/\max\{Price_1, \dots, Price_{T_k}\}$.  Next, we construct the user transactions network $G$ for $k$-th token on day $t$. From $G$, we calculate the  number of user transactions ${E}$. 

\medskip
\noindent\textbf{Model validity:} In our prediction models we use past  information up to and including day $t$ to predict anomalies for day $t+1$ (i.e., prediction horizon $h$ of 1) or days $t+h$ (i.e., longer prediction horizons $h$, $h>1$). Hence, these experimental settings ensure that no data leakage occurs.

\medskip
\noindent\textbf{Filtering.} Although Betti numbers provide a non-parametric solution to combine information on edge dissimilarity with node connectedness, the computational complexity of Betti calculations prohibits their usage in large networks. For example, for 2-simplicial complexes, \textquote{currently no upper bound better than a constant times $n^3$ is known}~\cite{edelsbrunner2014computational}. To decrease complexity, we induce a sub-network $G^\prime$ by selecting $K$ users who have the most edges in the network $G$. This filtering not only reduces the network size, but also removes network order fluctuations across time. Differences in Betti numbers of daily token networks can now be attributed to edges and their weights directly. From $G^\prime$, we then calculate 7-day rolling depth values~(\ref{RollDepth}) 
 $RD_7(\mathcal{B}_{0})$, $RD_7(\mathcal{B}_{1})$ and $RD_7(\mathcal{B}_{2})$, respectively. 
 
\begin{table}[]
    \centering
    \caption{Model descriptions}
    \begin{tabular}{ m{2.3em}  m{2cm} m{4.5cm}}
    \textbf{Model} & \textbf{Description} & \textbf{F: Model Inputs} \\ \hline
    $M1$    & {Baseline} & \makecell{$\mbox{PN}$, ${nE}$, $nV$, $GC$} \\ 
    \hline
    $M2$    & {Betti 0} & \makecell{$\mbox{PN}$, ${nE}$, $nV$, $GC$,  \\ $RD_7(\mathcal{B}_{0})$} \\  
    \hline
    $M3$     & Betti 0, 1& \makecell{$\mbox{PN}$, ${nE}$, $nV$, $GC$,  \\$RD_7(\mathcal{B}_{0}), RD_7(\mathcal{B}_{1})$}\\  \hline
    $M4$     & Full model& \makecell{$\mbox{PN}$, ${nE}$, $nV$, $GC$,  \\ $RD_7(\mathcal{B}_{0})$,$RD_7(\mathcal{B}_{1}),$  $RD_7(\mathcal{B}_{2})$}\\ \hline
    \end{tabular}%
\label{tab:model}
\end{table}

Rationale behind our modeling approach is that network topological features, summarized in terms of RD of Betti limits, add an important layer of information that can be missed by the traditional network summaries. Hence, to test the improvement in anomaly prediction due to adding the network topological features, we evaluate predictive performance of the four models listed in Table~\ref{tab:model}, using normalized token price 
($\mbox{PN}$), graph based (edge count ${nE}$, node count $nV$, average clustering coefficient $GC$) and topological variables (rolling depth values of Betti limits $RD_7(\mathcal{B}_{0}), RD_7(\mathcal{B}_{1}), RD_7(\mathcal{B}_{2}$)). Models are fitted using Random Forest (see Section~\ref{sec:expsetting}).

\section{Experimental Settings}
\label{sec:expsetting}
\begin{figure} 
\centering
\includegraphics[width=0.6\linewidth]{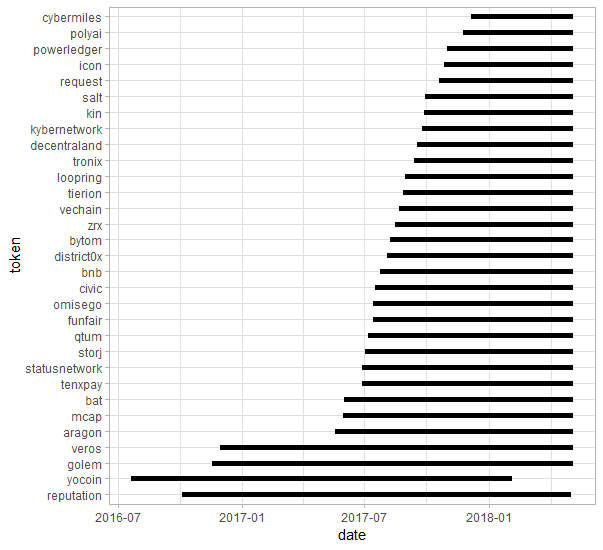}
\caption{Ethereum token start dates.
}
\label{fig:1erc20History}
\vspace*{-0.5cm}
\end{figure}
\textbf{Dataset.} We created our dataset by installing the official Ethereum Wallet and downloading all blocks. We used the EthR~(\url{github.com/BSDStudios/ethr}) library to query Ethereum blocks through the Go Ethereum Client (i.e., Geth). Our set contains all Ethereum data during 07/2015-05/2018, with a total of 5.5 million blocks. 
Our data and code are available at
\url{github.com/yitao416/EthereumCurves}.

By parsing the data, we discovered 1.7K ERC20 tokens which had more than 10K transactions. We included an ERC20 token in our analysis if it had  more than \$100M in market value, as reported by the \url{EtherScan.io} online explorer. This choice has resulted in 31 tokens and is motivated by a goal of developing {\it verifiable} prediction results on valuable tokens which likely will not fail and disappear in a short time.   On average, each token has a history of 297 days, with minimum and maximum of 151 and 576 days, respectively. The first dates of tokens on the Ethereum blockchain are reported in Figure~\ref{fig:1erc20History}.

\medskip
\noindent\textbf{Betti Descriptors.} We compute the Betti limits for up to $p=2$ (i.e., $\mathcal{B}_{0,.}$, $\mathcal{B}_{1,.}$ and $\mathcal{B}_{2,.}$) by using GUDHI (i.e., a generic open source C++ library for TDA~\cite{maria2014gudhi}). 
We a-priori set $K$ of 150 in the filtered network approach (see Section~\ref{sec:methanomaly}), as even for the most traded tokens such as Tronix and Bat, top 150 nodes in daily networks form 75\% and 80\% of all edges, respectively.
The filtered node approach effectively removes $20-25\%$ of edges in Betti calculations, which reduces computational costs.

\medskip
\noindent\textbf{Prediction Models.}
We set the first $2/3$ and $1/3$ of a token's timeline period as training and testing sets, respectively. We report our results based on Random Forest models which consistently outperform Box-Jenkins models for all prediction horizons. For example, at 2-day ahead forecasting, the best Box-Jenkins autoregressive integrated moving average (ARIMA) model with all predictors (M4) yields a prediction accuracy of 89\%, whereas our results for Random Forest model (M4) reach 94\%. 

For space limitations, we detail ARIMA settings and results in the supplementary material.
Each Random Forest model uses 500 trees, and sampling all rows of the dataset is done with replacement. 
Number of variables used at each split 
for all the four models is the floor of number of features. The models are implemented using
the \texttt{randomForest} package in \texttt{R}.

 Finally, for illustrative purposes we set the magnitude of the price shock $\delta=0.25$, following the guidelines
 on the trading cost perspectives
 by~\cite{claes2010stocks}. (For the detailed overview on the trader-defined choice of $\delta$ and associated investment strategies see~\cite{amini2013review, CFA2018, brady2019investors}.)  

\section{Experimental Results}
\label{sec:experiments}
We now illustrate what practical insights the resulting extracted
information on the local geometry and topology of the Ethereum transaction graph can bring into crypto-token analytics.

\subsection{Hidden Cointegration of Price and Graph Topology: How Do Tokens Co-move?}

Cointegration refers to a phenomenon when two economic or financial time series follow a common stochastic trend which is represented as a linear combination of system shocks~\cite{Engle:Granger:1991} -- that is, the two time series exhibit a similar response to shocks. In contrast, hidden cointegration analysis, as a variant of nonlinear cointegration, allows to assess a response of the two time series to various asymmetric system shocks, i.e. upward and downward movements due to, for example, positive and negative media news~\cite{granger2002hidden}.

To develop the best arbitrage trading strategy based on multiple assets~\cite{chan2013algorithmic}, the primary interest of many algorithmic trading platforms is to gain an insight on: which financial instruments exhibit joint co-movement trends?, and what can serve as a sign for future co-movement patterns? 
Intuitively, pairs of instruments that have exhibited co-movements in the past, are likelier to show co-movements in the future~\cite{malkiel1970efficient}.
Our study is then motivated by the following queries: Can cointegration in the currently observed local topological structures of crypto-tokens be a sign for future cointegration in crypto-token prices? Does this information contain an additional utility, compared to the cointegration of the currently observed crypto-token prices?

To address these queries, for each pair of tokens, we find their common trading time interval and equally divide it into two periods.
The hidden cointegration tests~\cite{Engle:Granger:1991, granger2002hidden} are then conducted in both periods for pairs of crypto-tokens in terms of their i) prices and ii) Betti descriptors. 

\begin{figure}[ht!]
 \centering
\begin{subfigure}{.25\textwidth}
 \includegraphics[width=0.9\linewidth]{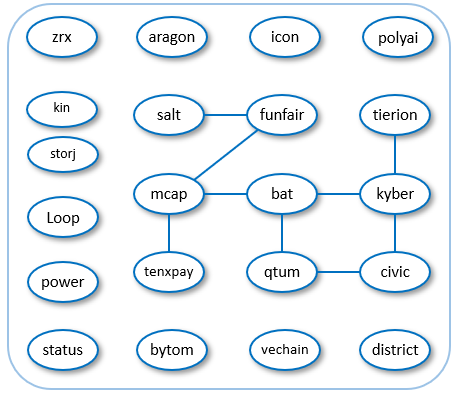}
  \caption{Price co-integration in tokens.}
  \label{fig:shocksPrice}
\end{subfigure}%
~
\begin{subfigure}{.25\textwidth}
   \includegraphics[width=0.9\linewidth]{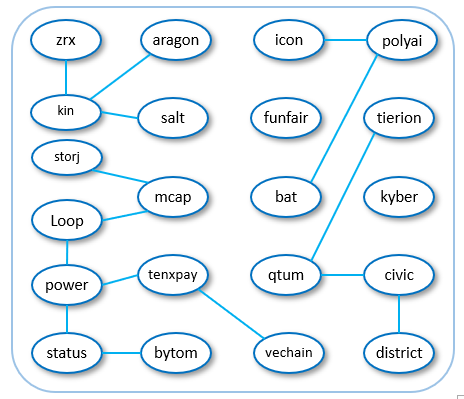}
  \caption{Betti co-integration in tokens.}
  \label{fig:shocksBetti}
\end{subfigure}

\caption{Cointegrated tokens that are also cointegrated in future price. An edge denotes cointegration.}
\label{fig:shocks}
\end{figure}

As Figure~\ref{fig:shocks} shows, only 9 pairs of crypto-tokens are cointegrated in price in both training and testing periods.  In contrast, in 15 cases a cointegration in Betti descriptors in the training period is also reflected in a crypto-token price cointegration in the testing period. 
Hence, we can conclude that \textit{previous cointegration in Betti descriptors of crypto-tokens might be a stronger sign for future cointegration in the prices of these crypto-tokens}.

Furthermore, price and Betti cointegrations found in the training period among the 31 considered tokens are almost disjoint, with the exception of the \textbf{civic}-\textbf{qtum} pair.
These findings suggest that \textit{local topology of crypto-token graphs is likely to contain important complementary information to more traditional data sources such as prices}.

\subsection{Performance in Crypto-Token Price Anomaly Forecasting}
\label{sec:exppredictability}

\begin{figure}[h]
    \centering
    \includegraphics[width=0.35\textwidth]{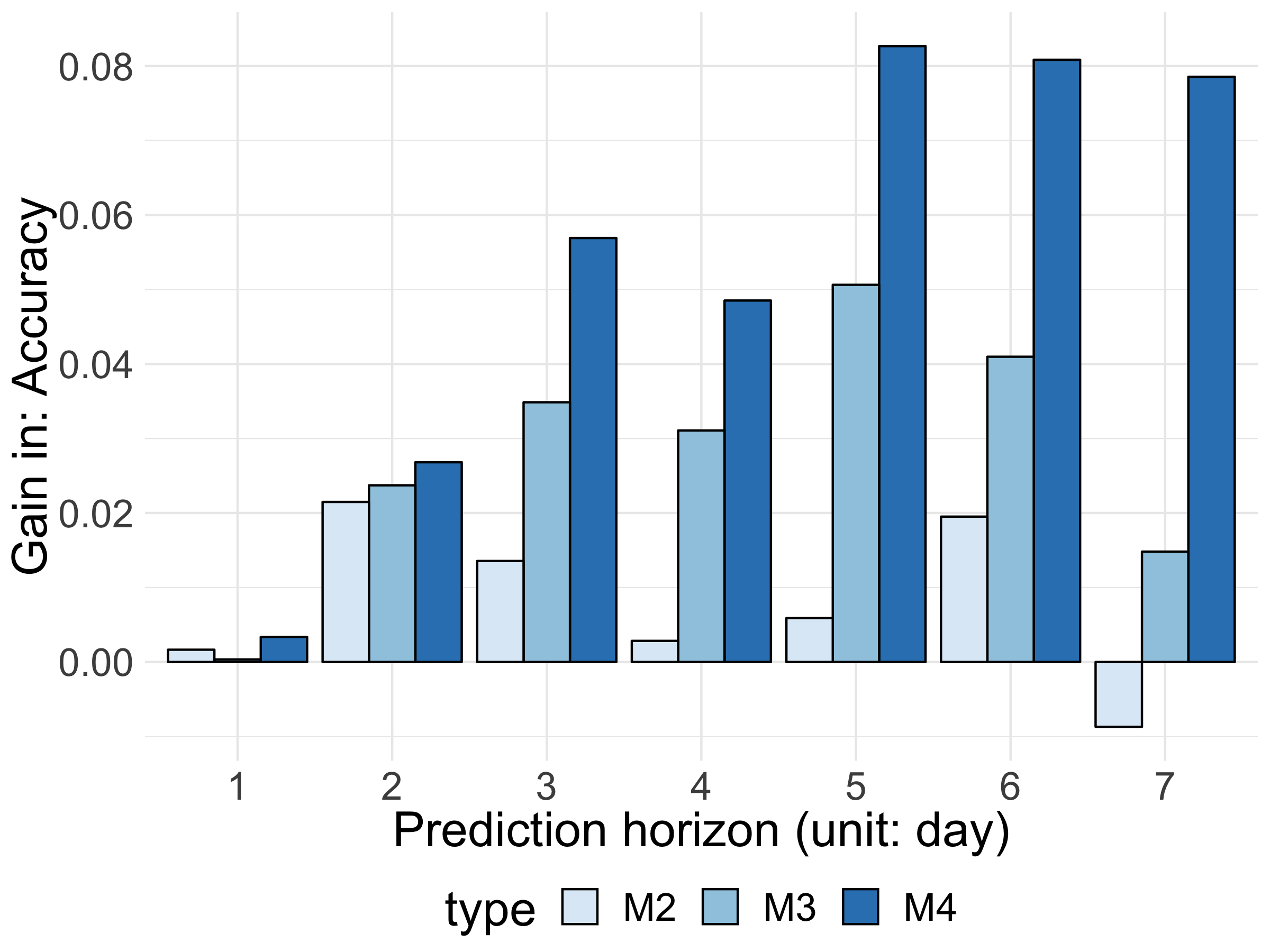}
    \caption{Model Accuracy}
    \label{fig:6accuracy}
\end{figure}

We predict price anomalies in 31 token networks, where a total of 9,042 days are predicted as anomalous (anomaly:true) or non-anomalous (anomaly:false).
On 145 of these days, a true price anomaly occurs, as defined by a change in the absolute price return of more than 0.25. Mean and median numbers of anomalies are 6.59 and 2 per token, respectively. The Veros token had a maximum of 46 anomalies. Nine tokens do not have any price anomalies in their test period (the last 1/3 of their timeline). In the days leading up to 2018 January, token prices exhibit substantial increases; on some days more than 20 tokens show $>0.25$ absolute price returns. In this period (Oct-Dec 2017) price of the Ethereum currency, ether, increased from \$305 to \$1,389. In 2018 Jan we see token prices decreasing sharply, but unlike the increase period, we observe fewer ($\leq 7$) anomalies in tokens on the same day.

\begin{figure}[h]
    \centering
    \includegraphics[width=0.28\textwidth]{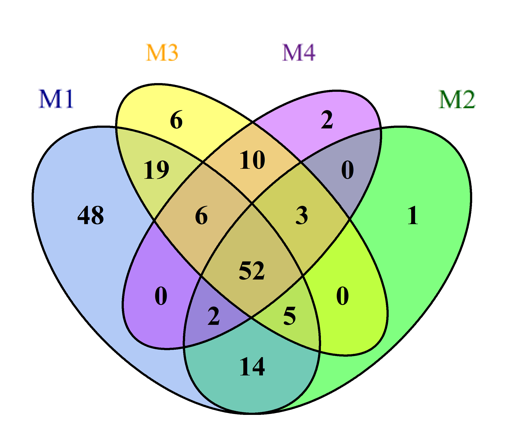}
    \caption{A Venn diagram for the number of predicted anomalies in all token networks for $h= 2$. Intersecting regions indicate agreement on predictions}
    \label{fig:9accuracydiagram}
\end{figure}

 Fig.~\ref{fig:9accuracydiagram} depicts the number of {anomaly/true} predictions by models. Models M2, M3 and M4 (Betti models) predict the same 138 days as anomalous. Additional 13  days are predicted anomalous by either only Betti model M2 (6 days) or Betti model M3 (7 days). Betti models make a lower number of anomaly:true predictions compared to the baseline M1 model, which uses traditional features such as price and number of edges. For example,  there are 186 true anomalies ($h=2$, i.e., anomalies in either of the next two days) in the considered token networks. M4 makes 146 anomaly/true predictions, and 86 of them are indeed true anomalies.  In M1 these values are 238 and 94, respectively. Compared to M4, M1 predicts 92 more days as anomalous, but only 8 of them are true anomalies.
 
Table~\ref{tab:tokenaccuracy} shows model accuracy values for the top ten tokens, ordered by average edge counts in daily networks. Models have high accuracy values, but for some tokens, such as icon, we reach high accuracy (i.e., 0.9) with the full model only.  We show the accuracy improvement over the baseline model M1 in Fig.~\ref{fig:6accuracy}. For up to 7-day horizons, all Betti models have a positive gain over the baseline model M1. Compared to other models, the M4 (full) model has the best performance as horizon increases from 1 to 7. The accuracy results offer evidence that Betti models are more conservative in making anomalous day predictions, and their accuracy is better than the baseline model M1.
 
\begin{figure}[t!]
 \centering
\begin{subfigure}{.40\textwidth}
 \includegraphics[width=1\linewidth]{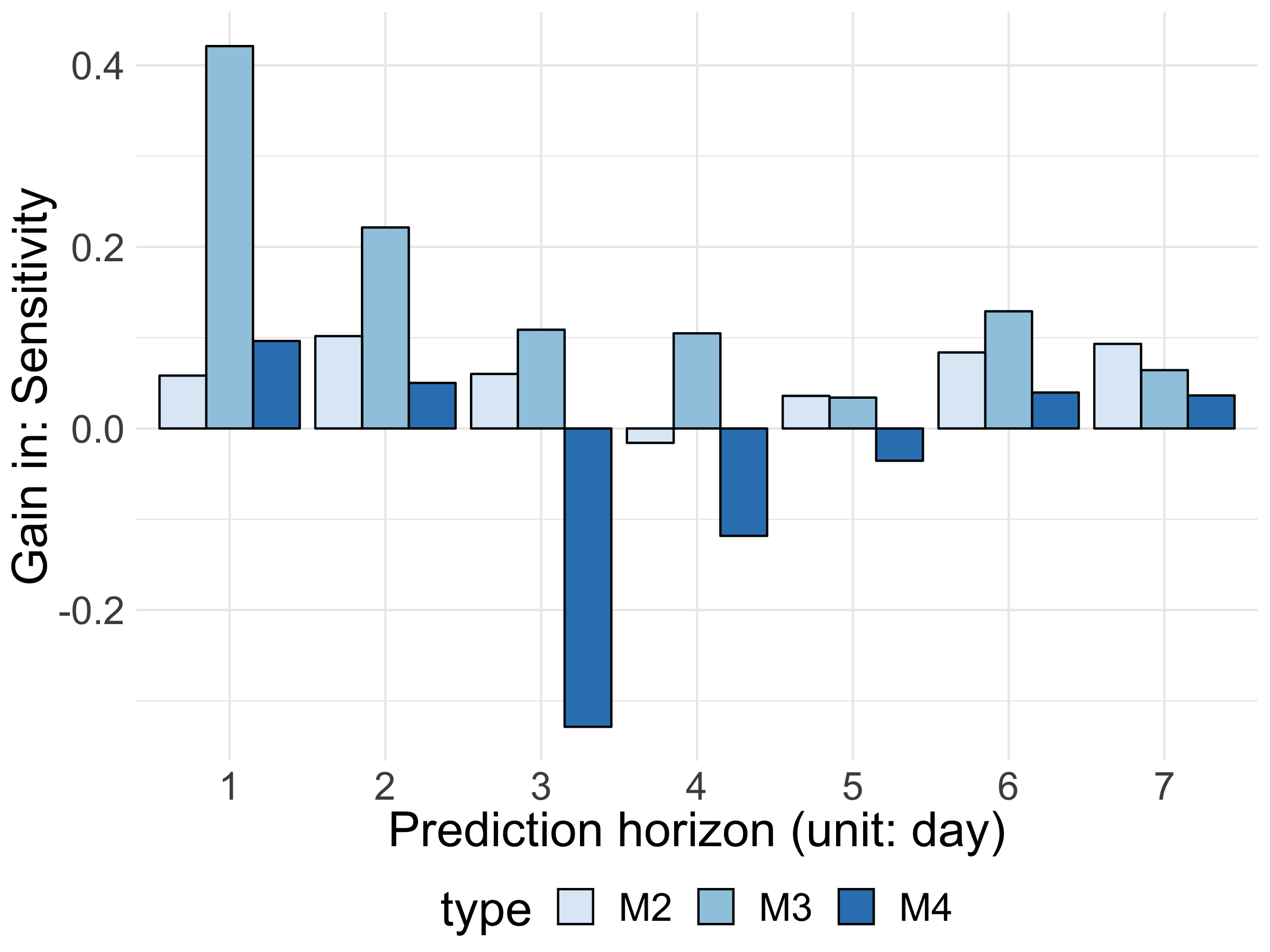}
  \caption{Recall (Sensitivity)}
  \label{fig:6sensitivity}
\end{subfigure}
~
\begin{subfigure}{.40\textwidth}
   \includegraphics[width=1\linewidth]{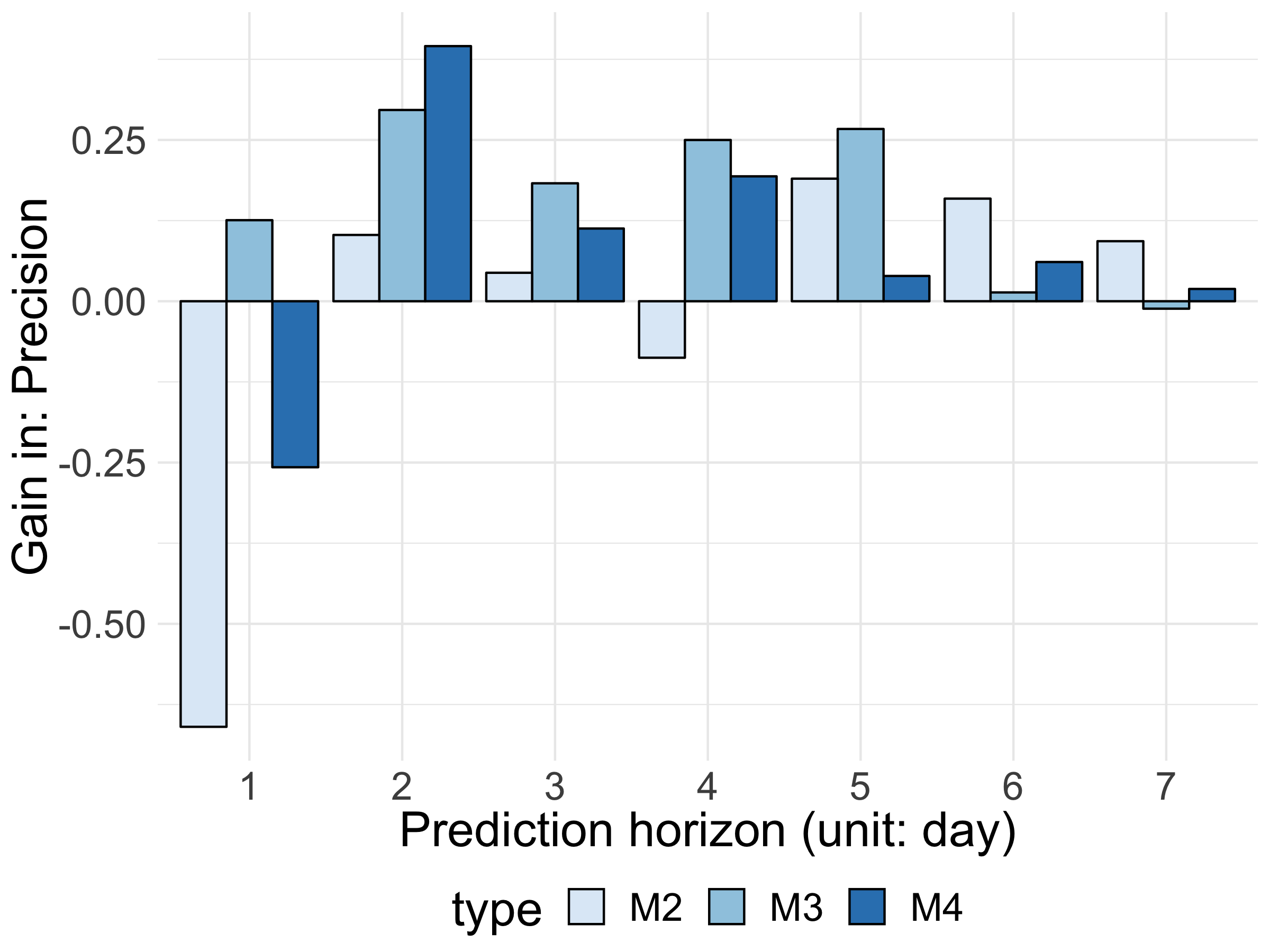}
  \caption{Precision}
  \label{fig:6precision}
\end{subfigure}
\caption{Performance for increasing horizon values.}
\label{fig:metrics}
\end{figure}

 The recall results in Fig.~\ref{fig:6sensitivity} show that M3 delivers the highest gain in recall for all horizons. Fig~\ref{fig:6precision} depicts the precision results.  For $h$ of 1, recall values are the highest but precision gains are negative. We achieve the best performance for $h$ of 2, where both precision (in M4) and recall (in M3) gains are over 20\%. 
 As M4 differs from M3 in its use of $\mathcal{B}_2$, we find the differing performance of M3 and M4 in Figures~\ref{fig:6sensitivity} and~\ref{fig:6precision} interesting. In particular, the results indicate that the use of $\mathcal{B}_2$ in M4 decreases the number of correctly predicted true anomalies, but increases the number of true anomalies in predictions.

Although predicting true negatives (non-anomalous days) is useful, the most important task of anomaly detection is to predict true anomalies well in advance. The unbalanced nature of our dataset complicates this task; only 1.58\% of all days are true anomalies, limiting the training cases to a few days per each token. For $h$ of 2, we achieve the highest average precision  of 0.393 per token in M4.

\begin{table}[]
    \centering
    \caption{Accuracy for $h = 2$ for the top-10 (by edge count $\bar{E}$) tokens.}
    \begin{tabular}{cccccc}\\\toprule  
\toprule
    \textbf{token} & \textbf{M1} & \textbf{M2} & \textbf{M3} & \textbf{M4} & \textbf{$\bar{E}$} \\
    \midrule
    tronix & 0.861 & 0.962 & 0.962 & 0.975 & 5198.2 \\
    omisego & 0.890 & 0.970 & 0.940 & 0.990 & 3027.7 \\
    mcap  & 0.887 & 0.904 & 0.913 & 0.913 & 1502.1 \\
    storj & 0.933 & 0.971 & 0.952 & 0.962 & 1224.3 \\
    bnb   & 0.927 & 0.969 & 0.979 & 0.979 & 1089.5 \\
    zrx   & 0.955 & 0.966 & 0.966 & 0.978 & 905.4 \\
    cybermiles & 0.922 & 0.961 & 0.961 & 0.961 & 872.7 \\
    vechain & 0.954 & 0.966 & 0.920 & 0.954 & 851.7 \\
    icon  & 0.754 & 0.877 & 0.877 & 0.908 & 783.5 \\
    bat   & 0.965 & 0.965 & 0.965 & 0.965 & 773.5 \\
    \bottomrule
\end{tabular}
    
    \label{tab:tokenaccuracy}
\end{table}

\section{Conclusions}

We have introduced the concepts of persistent homology and functional data depth to analysis of a yet untapped source of information on cryptocurrency dynamics: Ethereum transaction graph, and we have investigated such phenomena as price anomaly forecasting and hidden co-movement in pairs of tokens (Please see appendix for more details). Furthermore, we have proposed new functional summaries of topological descriptors, namely, Betti limits and Betti pivots. Our findings indicate that Betti pivots of the Ethereum transaction graph deliver up to 40\% improvement in precision over baseline methods in price anomaly prediction. Based on our analysis, we advocate that local geometry and topology of the transaction graph has a high utility in such important research directions on blockchain data analytics as health of the crypto-token ecosystem and identification of malicious trading activities. Furthermore,
the newly proposed concepts of Betti limits and Betti pivots and, more generally, a systematic linkage of TDA and FDA offer new perspective in data shape analysis way beyond blockchain applications. 

\section{Acknowledgments}

Gel has been partially supported by NSF DMS 1925346, IIS 1633331 and ECCS 1824716. Kantarcioglu has been partially supported by NIH 1R01HG006844, NSF CICI-1547324, and IIS-1633331.

\balance

\bibliographystyle{siam} 
\bibliography{curves}

\newpage
\appendix

\section{Supplementary Material}

This supplementary material consists of two parts. We provide the generation algorithm for Betti pivots and rolling depths (RD) on Betti limits, and discussion of performance of conventional time series models.
Symbols are listed in Table~\ref{tab:symbols}. 
 

\subsection{The Pivot generation algorithm}
 Algorithm~\ref{alg:numbers} details the generation process for Betti pivots and rolling depths (RD) on Betti limits.

\begin{figure}[h!]
   \centering
    \includegraphics[width=0.65\linewidth]{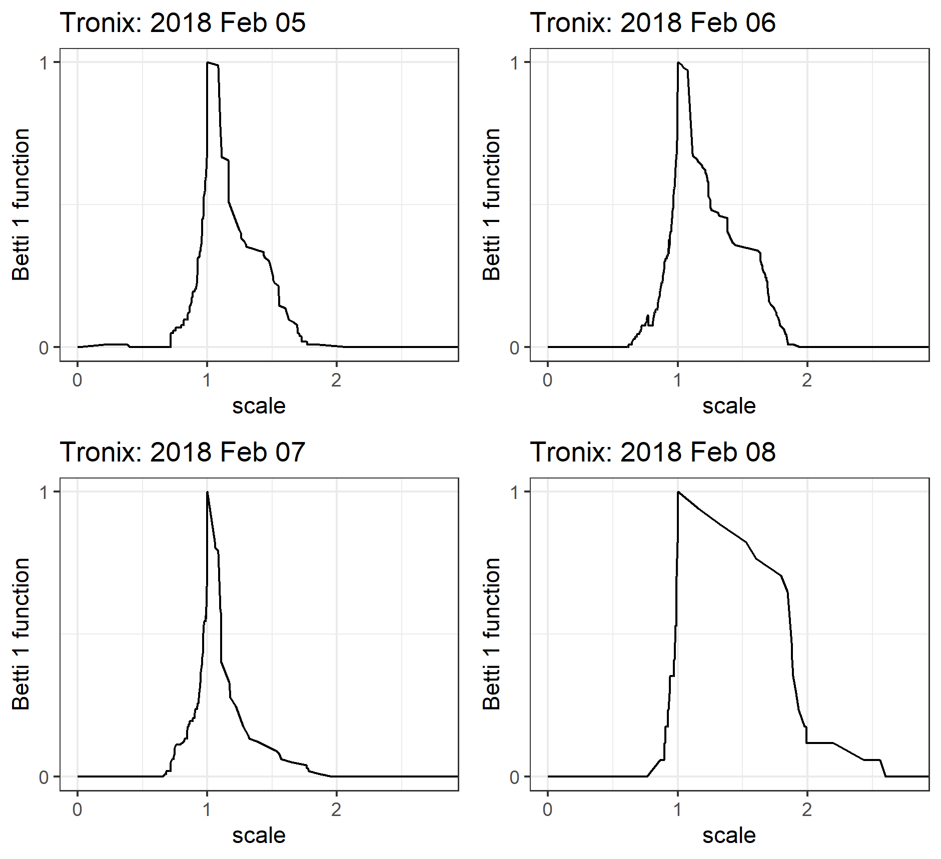}
    \caption{Betti numbers of the Tronix token. }
    \label{fig:4signature}
\end{figure}

Figure~\ref{fig:4signature} depicts discrete realizations of Betti limits (i.e., Betti numbers) for the Tronix token on 4 consecutive days in February 2018.

\subsection{Temporal Models}

To advocate the use of Random Forest classifiers, in this subsection we offer a glimpse into performance of traditional time series models based on conventional information sources as well as Betti pivots of the transaction graph. 

In time series analysis and forecasting, Autoregressive Integrated Moving Average (ARIMA) model with exogeneous regressors is a conventional benchmark    choice~\cite{brockwell2002introduction}. For each token, we divide the data set into training and test by ratio 2:1. The ARIMA model is constructed in the training set to predict anomalies in the test set. 

The optimal ARIMA model is selected based on the Hyndman-Khandakar algorithm~\cite{hyndman2007automatic} which considers unit root tests, the minimization of the corrected Akaike Information Criterion (AICc) and maximum likelihood estimator (MLE). We consider five models based on the employed features: the price autoregressive model and four dynamic regression models with different sets of lagged predictors. The lagged period is experimented from 1 day to 7 days; the lagged 3 days' predictors have the best price prediction. The calculation of prediction intervals is under the conventional assumption that the residuals are a white noise and follow a normal distribution. 

\begin{figure*}
    \centering
    \includegraphics[width=1\linewidth]{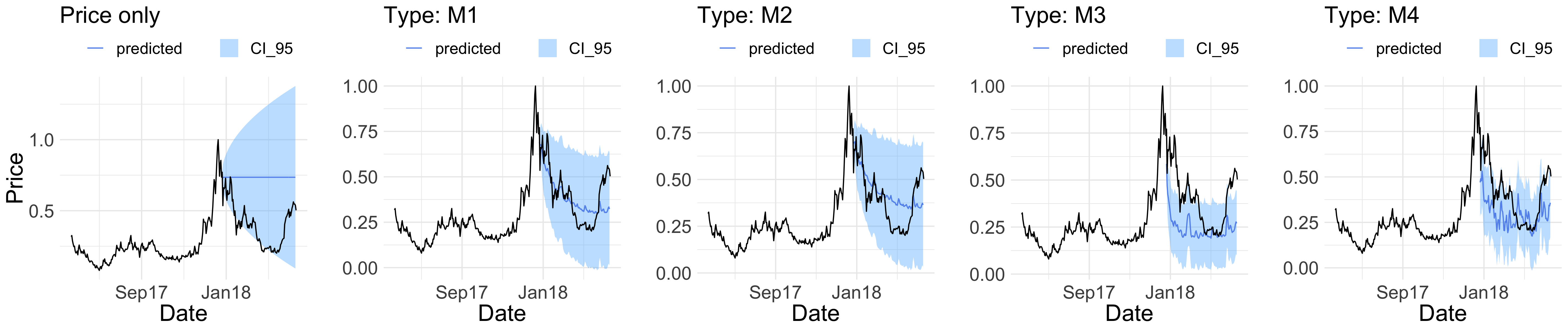}
    \caption{ARIMAx performance with topological models in the BAT token. The shaded region shows the vertical confidence interval ($\alpha=0.05$) around the predicted price. The M4 model contains all Betti predictors, and its predicted confidence interval gives the closest prediction to the actual price.}
    \label{fig:arimebat}
\end{figure*}

As Figure~\ref{fig:arimebat} shows, all dynamic regression models outperform the price autoregressive model. Compared with the benchmark (M1), the models with topological inputs have much narrow confidence intervals. Especially, for the shown Bat token, there exists a strong alignment between the full model prediction (M4) and the actual price movement.

\balance

\pagebreak
\begin{algorithm}[h] 
  \caption{Betti Pivots and Rolling Depths (RD) on Betti Limits Generation}\label{alg:numbers}
  \begin{algorithmic}[1]
    \Procedure{number}{$G$: token graph, $K$: filter, $d$: Betti dimension max, $w$: window} 
      \State induce graph $G^\prime$ for $top-K$ nodes
      \State compute $\tilde{\omega}_{uv}$ for each $ e=(u,v) \in G^\prime$ 
	 \For{Betti dimension p =\{0,\ldots,d\}} 
	 \For{each day $G^\prime_t \in G^\prime$} 
	 \State compute $\mathcal{B}_{p,t}$
	 \State 
	 {\footnotesize{$\mathcal{B}_p^s\gets \underset{\mathcal{B}_{p,t}\in\{\mathcal{B}_{p,t_1},\ldots,\mathcal{B}_{p,t_m}\}}{arg max}MBD(\mathcal{B}_{p,t}|\mathcal{B}_{p,t_1},\ldots,\mathcal{B}_{p,t_m})$}}
	 \State $F_p^t \gets RD_w(\mathcal{B}_{p,t})$
	 \EndFor
	 \EndFor	
	 \Return feature matrix $F$
    \EndProcedure
  \end{algorithmic}
\end{algorithm}

\begin{table} 
\caption{Symbols and notations.}\label{tab:symbols}
\centering
{
\begin{tabular}{cc}\\ 
Symbol& Explanation  \\\midrule
$\tilde{\omega}$ & extension of $\omega$\\
$F$ & feature matrix\\
$h$ & prediction horizon (in days)\\
$\delta$ & min price change for anomaly\\
$\epsilon$ & scale parameter\\
$\mathcal{C_\epsilon}$ & simplicial complex at scale $\epsilon$\\
$\mbox{R}_t$ & Price return for day $t$\\
$nE$, $nV$ & number of edges, nodes \\
$GC$ & average clustering coefficient \\
$\mbox{PN}$ & Normalized price\\
VR$_\epsilon$ & Vietoris-Rips complex at scale $\epsilon$\\
$\beta_p$,$\mathcal{B}_p$,$\mathcal{B}_p^s$ & Betti-$p$ number, limit and pivot\\
RD and MBD & rolling and modified band  depth\\
\bottomrule
\end{tabular}
}
 \end{table} 

\end{document}